\documentclass{article}

\usepackage[preprint,nonatbib]{neurips_2024} %submission is single blind
\usepackage[utf8]{inputenc} % allow utf-8 input
\usepackage[T1]{fontenc}    % use 8-bit T1 fonts
\usepackage{hyperref}       % hyperlinks
\usepackage{url}            % simple URL typesetting
\usepackage{booktabs}       % professional-quality tables
\usepackage{amsfonts}       % blackboard math symbols
\usepackage{nicefrac}       % compact symbols for 1/2, etc.
\usepackage{microtype}      % microtypography
\usepackage{xcolor}         % colors

\usepackage{graphicx}
\usepackage[numbers,sort&compress]{natbib}

\newcommand{\pquote}[1]{{``\textit{#1}''}}

\title{Dialogue with the Machine and Dialogue with the Art World: Evaluating Generative AI for Culturally-Situated Creativity}

\author{%
  Rida Qadri \\
  Google Research \\
    \And
  Piotr Mirowski \\
  Google DeepMind \\
  \And
  Aroussiak Gabriellan \\
  USC School of Architecture \\
    \AND
  Farbod Mehr \\
  Farbod Mehr Studio \\
    \And
  Huma Gupta \\
  Department of Architecture, MIT 
  \AND
  Pamela Karimi \\
  Cornell University \\
   \And
  Remi Denton \\
  Google Research \\
}

\begin{document}

\maketitle 

\begin{abstract}
This paper proposes dialogue as a novel and experimental qualitative method for evaluating generative AI tools for culturally-situated creative practice. Drawing on sociologist Howard Becker's concept of \emph{Art Worlds}, this method expands the scope of  traditional AI and creativity evaluations beyond benchmarks, user studies with crowd-workers, or focus groups conducted with artists. Our method involves two mutually informed dialogues: 1) `dialogues with art worlds,' placing artists in conversation with experts such as art historians, curators, and archivists, and 2) `dialogues with the machine,' facilitated through structured artist- and critic-led experimentation with state-of-the-art generative AI tools. We demonstrate the value of this method through a case study with artists and experts steeped in non-western art worlds, specifically the Persian Gulf. We trace how these dialogues help create culturally rich and situated forms of evaluation for representational possibilities of generative AI that mimic the reception of generative artwork in the broader art ecosystem. They also allow artists to shift their use of the tools to respond to their cultural and creative context. Our study can provide generative AI researchers an understanding of complex dynamics of technology, creativity and the socio-politics of art worlds, to build more inclusive machines for diverse art worlds.
\end{abstract}

\section{Introduction}

In his seminal work \emph{Art Worlds}, sociologist Howard Becker emphasized the fundamentally collaborative nature of art creation \cite{becker1976art}. Instead of focusing on a singular, isolated genius or on the aesthetic output, he argued for the study of an `art world' –  the network of critics, curators, suppliers, administrators, and audiences who shape artistic activity. The study of art in this conception should not be isolated to artistic outputs, but also encompass the social processes that make art production possible. Said differently, we should study what artists do and how they do it, not just artistic symbolism or composition. Art production in this view is  an inherently social process, not an individual or technical one. This scholarship changed the way art was studied \cite{becker2023art}.

As generative AI is integrated into the artistic domain, it also enters into existing art worlds and creates the possibility of shaping its own art worlds \cite{manovich2024artificial}.  Applying Becker's perspective to generative AI, the existing evaluation paradigm  of AI for creativity must necessarily become `social' and thus expand  in both its focus (beyond the output) and in who it involves (beyond the artist).  Yet, existing benchmark-centered studies on creativity have focused on quantitative metrics for outputs (e.g., based on human evaluations of the generated image's quality vs. its prompt such as in DrawBridge \cite{saharia2022photorealistic}, PartiPrompts \cite{yu2022scaling}, or conversely, of the ambiguity of crowdsourced annotations of images \cite{wang2023computational}) or on automated methods and datasets for Artistic Image Aesthetic Assessment \cite{yi2023towards}, including measures of the diversity of generated images \cite{ibarrola2024measuring}, of their novelty vs. an existing dataset \cite{jha2022creative} or conversely, similarity to an desired image style \cite{gallego2022personalizing}.  

More recently, there has been an emerging focus on socio-technical evaluations of AI. These approaches draw on qualitative methods, bringing in more participatory perspectives, and ultimately expand the frame of what is being evaluated \cite{wingstrom2024redefining,johnston2024understanding,caramiaux2022explorers,qadri2023ai,nordstrom2023evolving,shi2023understanding,bird2024artists,rajcic2024towards,shelby2024generative,srinivasan2021biases}. While these get closer to understanding AI as a socio-technical artifact, these studies  have primarily centered on conversations with artists through interviews, with only a handful studies focusing on artists experimenting with tools directly or bringing artists into conversation with one  \cite{rajcic2024towards,shelby2024generative,mirowski2023co,chakrabarty2023creativity,huang2023inspo,ippolito2022creative}. Their scope is also often narrowly focused  on the usefulness of a generative AI tool for artists as a  \emph{creativity support tool} \cite{cherry2014quantifying,shneiderman2007creativity}, even as they generate great insight into the individual artist's interactions with AI and demonstrate the various ways in which artists use and conceptualize emerging AI tools. Other qualitative studies focus on economic impacts for artists \cite{kawakami2024impact,epstein2023art,jiang2023ai} and perceptions of creativity and collaboration \cite{wingstrom2024redefining,johnston2024understanding}. The scope and method of these qualitative studies could be expanded to explore the ways in which generative art will be received by art worlds, shape new art worlds, or how collectively cultural communities could push the possibilities of representation in generative AI.

We fill this gap by introducing an experimental dialogic evaluation of AI for creativity coupling two types of dialogues: dialogue with the machine and dialogue with the art world. These mutually informed dialogues are a step towards a more culturally informed evaluation of AI in art, that can mimic its reception in the art world,  internalize its social impacts while also honing in on the particular creative processes it enables or constrains.  

In this paper we present the application of our approach within a concrete case study where we explored generative AI as a creative tool for cultural representation \cite{mohamed2020decolonial}, within the cultural context of the Persian Gulf. We instantiated the dialogue with machines as a multi-week experimentation period with artists using state-of-the-art generative AI tools. We instantiated the dialogue with the art world as workshops and reflections with artists and other stakeholders of the collective art world that AI is entering into, such as art historians, curators and archivists. The dialogue was mediated by researchers working in an AI lab who acted as facilitators of the workshops, by asking questions and pairing artists and experts for 1:1 dialogue. The remainder of the paper is structured as follows. First, we introduce our dialogic evaluation method, motivating its value and detailing how we approached the two dialogues within our case study. Next, we  present our case study and trace two dialogues that exemplify the value of this approach. Finally, we close by reflecting on our approach and the unique insights offered for Generative AI development through dialogue as a method. As a demonstration of the value of elevating diverse forms of multi-disciplinary expertise for AI evaluations, this paper has been co-authored with scholars and artists who participated in the study.

\section{Dialogue as Method}

In this section we introduce two mutually informed sets of dialogue: dialogue with art worlds and dialogue with machines. Throughout this section, we motivate the value of each dialogue and then describe our approach to instantiating the dialogue within our case study situated within the cultural context of the Persian Gulf. 

\subsection{Dialogue with Art Worlds: Evaluating Creativity as an Ecosystem}

According to Martin Zeilinger, human creativity is ``dialogic, relational, and fundamentally intertextual''  – a social act \cite{zeilinger2021}. One of the lessons of art history is that art was not developed in vacuum, but has been a dialogue between creator, receivers, audience, about symbols, processes and interpretation \cite{erwin1955meaning,hauser1951social,becker1976art}. Then, evaluations  of art are also necessarily social. We propose a method that allows us to engage the creative ecosystem to understand the impacts of AI in art production. By engaging experts in existing art worlds, we aim to expand the idea of who can evaluate generative AI, arguing that these evaluations should not just be the remit of AI researchers or of atomic communities of artists. 

To construct a dialogue with art worlds,  we placed artists in conversation with a multi-perspective panel of cultural experts (hereafter referred to as “commentators”) to evaluate different dimensions of generative AI, such as how generative AI could shape cultural relevance and meaning of artworks, the forms of agency it affords or that artists can wrest from it and its possibilities as a tool for cultural and political representation. By applying these critical reflections on the generative AI production process  and the outputs artists produced during the study, this process mimics what the reception of the generated artwork might be. 
 
We set up two interactive workshops. One workshop preceded the experimentation period of artists (discussed in Section \ref{sec:dialogue_with_machines}) and one workshop followed it. Before the workshops we also asked each participant to record an introductory video showcasing their work and ideas for the study, which were viewed by all participants before the workshop. During workshop 1, we collectively reflected on the possibilities of generative AI broadly and historical intersections of technology and art. Commentators and artists also broke up into one-on-one loosely structured conversations to collectively develop ideas for the artists’ projects. In the second workshop we collectively evaluated artists' processes and outputs, situating them within broader histories of art in the cultural context of the artists. Artists also had opportunities to attend `office hours' with commentators and researchers and to discuss their projects throughout the experimentation period. 

\begin{figure}
    \centering
    \includegraphics[width=1\textwidth]{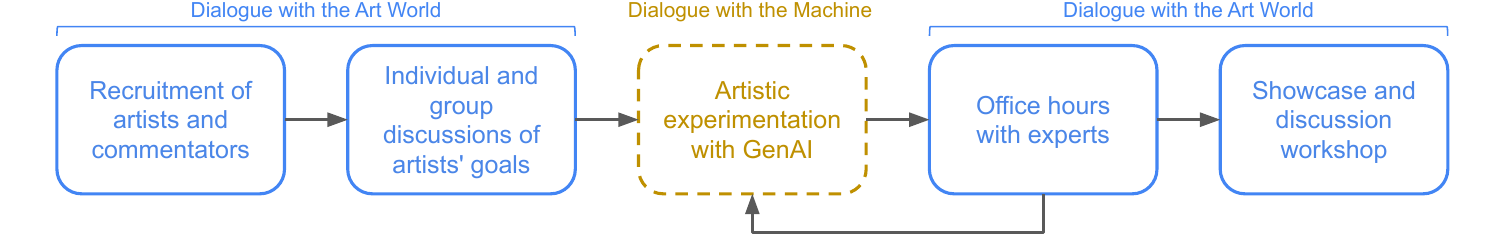}
    \caption{Overview of the dialogue-based approach followed in this study.}
    \label{fig:dialogue}
    \vspace*{-0.5cm}
\end{figure}

\subsection{Dialogue with Machines: Structured Artistic Experimentation with Generative AI Tools}
\label{sec:dialogue_with_machines}

AI researchers have called for an investigation into ``how the novel AI technologies are used in everyday creative processes'' \cite{wingstrom2024redefining} which can  shift AI towards ``more diverse forms of visual art collaboration and assistance'' \cite{johnston2024understanding}. Creating room for free form experimentation by artists during our study allows us to mimic real world use similarly to workshops done in \cite{rajcic2024towards}. However, one key difference in our study is that experimentation with the machine is informed by the dialogue with the commentators, to mimic the reception of generated art in the art world. 

Within our study, artists participated in a multi-week experimental study where they used generative AI tools to produce culturally situated media artifacts that tell stories of their own identities that have traditionally not been represented in dominant media.  Artists were free to use any model\footnote{Image generation tools used by artists included, among others, MidJourney \url{https://www.midjourney.com/home}, DALL-E \url{https://labs.openai.com/}, Adobe Firefly \url{https://www.adobe.com/products/firefly.html}, Stable Diffusion running on Replicate \url{https://replicate.com/} or on Runway ML servers \url{https://runwayml.com/}, and Google Colab \url{https://colab.research.google.com/}.}, technique, and dataset they wanted. Ideas for what they explored and created during this period were informed by the dialogue with commentators in the first workshop, and shaped through periodic meetings with commentators throughout the experimentation period. Artists were encouraged to experiment with prompt engineering and fine-tuning as approaches to steer models towards their desired representations. Artists documented and shared their process and experience in a series of process logs and reflection videos. 

This method allows us to study not just artistic outputs but also processes of artistic production. The study of such processes is significant in extending the artists' use of generative AI tools beyond the final artistic product, and also its impacts on the way they produce things. The process is as important in art as the output. Knowing about creative processes can also help us identify gaps in design, interface, interactions of generative AI systems and thus present developers with actionable pathways for improving generative AI as a tool for cultural representation. Moreover, having this process be informed by collective reflections also helps us see how far artists can push existing tools for more radical possibilities. This means insights and recommendations that emerge from this mutually informed dialogue do not just improve the interface of the technology but also its possibilities of use in different cultural contexts and use cases. 

\section{Case Study: Art Worlds of the Persian Gulf}
We focused our project on artists from a similar cultural background since we were interested in fostering a dialogue on generative AI as a tool for cultural representation. All artists and expert commentators had expertise in the Islamic worlds broadly and the Persian gulf specifically. 

\subsubsection{Participant recruitment} 
We recruited three artists and three commentators through personal networks of professional collaboration. Prior work with art worlds \cite{qadri2023ai} had helped us build connections that we leveraged for a snowball sampling of artists who both 1) had some experience with digital computational tools for creativity in their artistic practice and 2) identified with the Persian Gulf region (either through lived experience, heritage, or through their work). We similarly recruited art commentators who had expertise on the art world and social histories in the Persian Gulf.

The artists' expertise include graphical art, video, sculpture, architecture and design, and their work has been exhibited, among others, at Saatchi or Christie's galleries, at the Venice Biennale or at Centre Pompidou. The commentators hold positions at universities in the US and museums in the UK similar to the British Museum, Tate and V\&A. All commentators and artists were deeply immersed in the art worlds of the Persian Gulf. Two artists were Iranian with extensive history of making art representing Iranian contexts, one artist was of Armenian descent (a minority group within Iran) and a landscape scholar, one commentator was an expert in Iranian art history and media,  one commentator had training in the architecture and history of Islamic worlds and one commentator was the Iran curator at a major global museum in the UK. This sampling ensured contextually focused conversations. 

The artists’ and commentators’ names are withheld for privacy. Throughout the following subsections we refer to artists with a numeric participant identifier (e.g. Artist-1, Artist-2, and Artist-3). We refer to commentators based on their expertise (e.g. Art Historian, Architectural Historian, Curator). 

\subsection{Tracing the Benefit of Dialogues}

% In this section we present evidence of the contribution this method offers by tracing two dialogues that helped distill progressive and actionable design recommendations (Dialogue A) and evaluate the representational possibilities of generative AI (Dialogue B). 
In this section we present evidence of the utility of this method for technology developers by tracing two dialogues that i) distill decentralized datasets as a pathway for improved cultural representation (Dialogue A) and ii) surface possibilities for culturally situated generative AI use-cases within creative domains (Dialogue B). Coupling the the artists' experimentation with the dialogues with commentators allowed us to interrogate and understand what culturally situated creativity is possible through the current technological state, and what interventions might be needed to give more agency to non-western artists. The collective conversation particularly helped us learn from histories of past technologies intervening in artistic practice, to gain an understanding of how we can analyze generative AI as a technology of artistic practice. These dialogues also allowed artists and commentators to collectively develop  visions for generative AI through their experimentation, practice, and discussion that were suitable to the art worlds they were part of and attuned to the different processes of the artists.

\subsubsection{Dialogue A: Aspirations for decentralized datasets}

We present a dialogue where multiple artists developed a specific pathway towards more culturally situated generative AI: building decentralized datasets. We present the multiple dialogues between the artists and commentators  and artists and machines, through which artists iteratively added layers of specificity for this pathway.

One exchange that led to this aspiration was between Artist-3 and the Museum Curator, discussing the limitations of existing datasets for female Iranian representation such as Qajar\footnote{Qajar art corresponds to the architecture and paintings produced during the Qajar dynasty, which lasted from 1781 to 1925 in Persia / Iran.} paintings or museum datasets. To address these limitations, Artist-3 developed an innovative approach by creating three distinct AI models, each drawing from different cultural sources. The first model utilized Instagram photos to capture contemporary representations of Iranian women, reflecting the modern digital landscape. This approach prompted the Museum Curator and Artist-3 to collectively explore the exclusions embedded in social media platforms, which may reflect certain normative standards. Artist-3 understood the need to counterbalance these modern influences in visual data with more grounded, historical perspectives. The second model therefore delved into historical depth by analyzing vintage photos from what the Museum Curator characterized as a \pquote{counter archive}: the Qajar Women's World Archive\footnote{\url{http://www.qajarwomen.org/en/}}, which presented the female perspective on Qajar dynasties. The third model was trained on traditional Iranian paintings, introducing an artistic layer that rooted the character in cultural heritage. These three models were looped together by the artist in an iterative process to create a visual character that  embodied a mosaic of lived experiences, historical narratives, and cultural artistry. This dialogue between the machine, artist and commentators ultimately prompted the suggestion of a \pquote{decentralized library,} akin to Wikipedia, to drive AI datasets towards a more collective effort which could push models towards more culturally rich forms of visuality: \pquote{decentralized and everyone are inputting and people are saying when there is some part missing, they can go and find and add to it and if it's wrong, other people can edit.}

Another dialogue, between Artist-1 and the Art Historian, also focused on decentralized datasets. However, this dialogue resisted the idea that openness was an inherent good, and instead suggested decentralized libraries with restricted access as means of protecting community knowledge. This exchange started with a more theoretical concern expressed by the Art Historian.  In response to suggestions of more data collection to improve representation of different global cultures within generative AI, the Art Historian  expressed a feeling of discomfort about AI being able to know everything about Iran: \pquote{but I personally feel like maybe it's OK for Iran to have these secrets. Why do we need to let technology to know this complex, complicated country through and through?} This provocative prompt evolved into ideas around decentralized datasets with access restrictions to enable \pquote{protecting knowledge while giving access to knowledge} (Artist-1). Artist-1  reflected on the need for systems \pquote{built for us by us… Or by us for us} and in conversation with the Art Historian reflected on how they could \pquote{actually practice that.} They suggested, as an example, a dataset that only \pquote{specific people could use based on a specific archive, which was not always available to everyone,} imagining systems that can be localized and decentralized by \pquote{building your own library, pulling something, making something that data doesn't go anywhere?} For the Art Historian, this was a way to decolonize technology \cite{mohamed2020decolonial} by making sure data would not \pquote{be used for the same power structures that create all the sources of knowledge... not necessarily producing it for the same institutions that marginalized those people in the first place.}

So while Artist-3’s idea of decentralized libraries imagined these datasets as a sort of collective commons, the Art Historian and Artist-1’s reflections questioned the inherent value of open access and inclusivity without setting the terms of inclusion: \pquote{is inclusivity, is open source inherently good, you know, which the answer is no, it's not inherently good} (Artist-1). This led Artist-1 to approach the idea of decentralized libraries from the perspective of \pquote{protection of knowledge systems.}  Both these exchanges demonstrated how iterative exchanges with commentators and the machine can help artists develop more novel, progressive, and recommendations for more inclusive generative AI. The unique pathways both conversations took showed how the same aspiration can have different instantiations and operationalization.
 
\subsubsection{Dialogue B: Using Generative AI for  Representational Possibilities}

In this section we trace another dialogue--between one artist, multiple commentators, and the machine--that shaped ideas of how generative AI output could be used for cultural representation. We show how investigating, through dialogue, the historical context of an artwork, the intention of the artist, and their knowledge about the cultural context within which they are working offers a new lens into the art itself and can even push the artist to think deeper about their work.  

This exchange began with a discussion between one artist and the commentators in workshop 1, that ultimately shifted the artist’s project idea for the experimentation period. Artist-2 was struck by two highly divergent understandings of Persianness articulated by two different commentators. The first focused on a historical outsider’s view of Persianness, framed through \pquote{the Western Cultural imaginary… through certain kinds of artifacts, through a particular period in history} (Museum Curator). The second offered an \pquote{on the ground} understanding of Persianness defined by \pquote{everyday life through everyday forms of resistance} (Art Historian). Inspired by this, Artist-2 used their project to push against the fixity of definitions that AI might produce to shape a particular idea of Persianness. They explored this idea using generative AI to create what they referred to as \pquote{hybridized} imagery that integrated slogans known to represent acts of resistance (such as ``Woman, Life, Freedom'') into historical cultural artifacts (such as carpets and ceramics). Ultimately, their goal was to leverage generative AI to \pquote{reimagin[e] these historical artifacts as new sites for activism} (Artist-2).

In workshop 2, the artists and commentators reconvened to collectively reflect upon the generated images and the way generative AI offered up a potentially new avenue for responding to colonial histories and outsider’s views of Persianness.  One commentator reflected on the inherently political history of carpets, and was intrigued by the Artist-2's intent to explicitly politicize carpets through the new synthetic and hybridized carpet images.  Moreover, Artist-2 was aesthetically pleased with the seamless integration of the graffiti into the carpets and ceramics: \pquote{the graffiti look like threads of the carpet were part of the graffiti that was literally woven into it. And same with the graffiti so that it felt like it was embedded within the design of the polychrome ceramic kind of patterning}. At the same time, the group reflection on the synthetic images also surfaced a nuanced critique of current generative AI capabilities. While the generated images integrated the slogans into the carpets and ceramics in aesthetically compelling ways, the text rendered was \pquote{gibberish} and \pquote{wasn't communicating anything language-wise} (Artist-2). While participants recognized that text generation capabilities are limited across the board in current generative AI tools, the stakes of this particular failure were high. 
Commentators drew the connection between the generated images and cultural appropriation through pseudo-calligraphy utilized during the Byzantine and Renaissance periods: \pquote{this reminds me of 19th century orientalist paintings when they paint carpets and they go squiggle, squiggle, squiggle} (Museum Curator). Despite these representational limits, commentators and artists noted the role of art as social commentary, and Artist-2 reflect on how they would want an audience to interact with such art: \pquote{if it's generating something that's pointing to a period and kind of continues the same kind of historical patterns, then I would want to work with that, but kind of point it against itself. So it would still be an activist, political act.} This exchange also convinced commentators that generative AI could have a more \pquote{radical potential} for cultural representation. The pushing of these representational possibilities were high stakes; Architectural Historian noted the risk of AI \pquote{flattening and erasing, and pushing our imagination towards a much narrower range.}

These exchanges shifted the artist’s use of generative AI tools to explore radical possibilities situated within the cultural ecosystem of the artist and the broader role of art in society. Experimentation showed how artists might push the tools forward in light of political histories and use generative AI as a tool for intervening in those histories or for offering critical commentary on those histories. These exchanges also offered a nuanced evaluation and critique of generative AI limitations, contextualized within colonial histories and outsiders' views of Persian culture and art that could provide important takeaways for the design of generative AI tools that do not repeat the mistakes of the past while also giving inspiration for methods of interrogation possibilities of generative art for artists. Here, the idea of what artists may want to explore and evaluation of the output came through dialogue with commentators. 

\section{Discussion and Takeaways}

The dialogic method we propose allows us to evaluate AI's role within a culturally-situated ecosystem of creativity in a particular context. We were able to understand elements of generative AI’s outputs which would be evaluated within their reception in the art world. Our approach also recognizes that there are more impacted communities in this ecosystem that have essential expertise for appropriate evaluation of models within particular cultural context and use cases. By elevating their expertise we follow other forms of community-centered evaluations of AI \cite{mirowski2024robot,qadri2023ai,gadiraju2023wouldn,mack2024they} and add to existing methods for evaluating the complex interplay between technology, creativity and socio-politics of the art world. Such holisitic engagement can offer  metrics for AI creativity tools that are more relevant to the non-western world, fostering a more nuanced discourse on the role of AI in creative processes. 

Our dialogic approach surfaces design pathways that are  attuned to the broader socio-political ecosystem of creativity and culture and can help build tools which could live up to art’s representational and social potentials, especially in non-western contexts. We did not have space to dive into the aspirations and recommendations that emerged from this study, which will be published separately as a full paper. Nonetheless, we offered up examples of mutually-informed dialogues between artists, commentators, and machines to showcase how this approach shifted artists’ use of the tools to test for more radical possibilities rooted in their cultural context and how they developed pathways that could drive the development of culturally relevant and meaningful tools for non-western artists.

\bibliographystyle{unsrt}
\bibliography{main}

\begin{thebibliography}{10}

\bibitem{becker1976art}
Howard~S Becker.
\newblock Art worlds and social types.
\newblock {\em American behavioral scientist}, 19(6):703--718, 1976.

\bibitem{becker2023art}
Howard~S Becker.
\newblock {\em Art worlds: updated and expanded}.
\newblock Univ of California Press, 2023.

\bibitem{manovich2024artificial}
Lev Manovich and Emanuelle Arielli.
\newblock {\em Artificial Aesthetics: Generative AI, Art and Visual Media}.
\newblock \url{https://manovich.net/index.php/projects/artificial-aesthetics},
  2024.

\bibitem{saharia2022photorealistic}
Chitwan Saharia, William Chan, Saurabh Saxena, Lala Li, Jay Whang, Remi Denton,
  Kamyar Ghasemipour, Raphael Gontijo~Lopes, Burcu Karagol~Ayan, Tim Salimans,
  et~al.
\newblock Photorealistic text-to-image diffusion models with deep language
  understanding.
\newblock {\em Advances in neural information processing systems},
  35:36479--36494, 2022.

\bibitem{yu2022scaling}
Jiahui Yu, Yuanzhong Xu, Jing~Yu Koh, Thang Luong, Gunjan Baid, Zirui Wang,
  Vijay Vasudevan, Alexander Ku, Yinfei Yang, Burcu~Karagol Ayan, et~al.
\newblock Scaling autoregressive models for content-rich text-to-image
  generation.
\newblock {\em arXiv preprint arXiv:2206.10789}, 2(3):5, 2022.

\bibitem{wang2023computational}
Xi~Wang, Zoya Bylinskii, Aaron Hertzmann, and Robert Pepperell.
\newblock A computational approach to studying aesthetic judgments of ambiguous
  artworks.
\newblock {\em Psychology of Aesthetics, Creativity, and the Arts}, 2023.

\bibitem{yi2023towards}
Ran Yi, Haoyuan Tian, Zhihao Gu, Yu-Kun Lai, and Paul~L Rosin.
\newblock Towards artistic image aesthetics assessment: a large-scale dataset
  and a new method.
\newblock In {\em Proceedings of the IEEE/CVF Conference on Computer Vision and
  Pattern Recognition}, pages 22388--22397, 2023.

\bibitem{ibarrola2024measuring}
Francisco Ibarrola and Kazjon Grace.
\newblock Measuring diversity in co-creative image generation.
\newblock {\em arXiv preprint arXiv:2403.13826}, 2024.

\bibitem{jha2022creative}
Divyansh Jha, Kai Yi, Ivan Skorokhodov, and Mohamed Elhoseiny.
\newblock Creative walk adversarial networks: Novel art generation with
  probabilistic random walk deviation from style norms.
\newblock In {\em ICCC}, pages 195--204, 2022.

\bibitem{gallego2022personalizing}
Victor Gallego.
\newblock Personalizing text-to-image generation via aesthetic gradients.
\newblock {\em arXiv preprint arXiv:2209.12330}, 2022.

\bibitem{wingstrom2024redefining}
Roosa Wingstr{\"o}m, Johanna Hautala, and Riina Lundman.
\newblock Redefining creativity in the era of ai? perspectives of computer
  scientists and new media artists.
\newblock {\em Creativity Research Journal}, 36(2):177--193, 2024.

\bibitem{johnston2024understanding}
Hannah Johnston and David Thue.
\newblock Understanding visual artists’ values and attitudes towards
  collaboration, technology, and ai.
\newblock In {\em Graphics Interface 2024 Second Deadline}.

\bibitem{caramiaux2022explorers}
Baptiste Caramiaux and Sarah Fdili~Alaoui.
\newblock " explorers of unknown planets" practices and politics of artificial
  intelligence in visual arts.
\newblock {\em Proceedings of the ACM on Human-Computer Interaction},
  6(CSCW2):1--24, 2022.

\bibitem{qadri2023ai}
Rida Qadri, Renee Shelby, Cynthia~L Bennett, and Remi Denton.
\newblock Ai’s regimes of representation: A community-centered study of
  text-to-image models in south asia.
\newblock In {\em Proceedings of the 2023 ACM Conference on Fairness,
  Accountability, and Transparency}, pages 506--517, 2023.

\bibitem{nordstrom2023evolving}
Paulina Nordstr{\"o}m, Riina Lundman, and Johanna Hautala.
\newblock Evolving coagency between artists and ai in the spatial cocreative
  process of artmaking.
\newblock {\em Annals of the American Association of Geographers},
  113(9):2203--2218, 2023.

\bibitem{shi2023understanding}
Jingyu Shi, Rahul Jain, Runlin Duan, and Karthik Ramani.
\newblock Understanding generative ai in art: An interview study with artists
  on g-ai from an hci perspective.
\newblock {\em arXiv preprint arXiv:2310.13149}, 2023.

\bibitem{bird2024artists}
Charlotte Bird.
\newblock Artists and ai: Creative interactions and tensions.
\newblock In {\em Extended Abstracts of the CHI Conference on Human Factors in
  Computing Systems}, pages 1--6, 2024.

\bibitem{rajcic2024towards}
Nina Rajcic, Maria~Teresa Llano~Rodriguez, and Jon McCormack.
\newblock Towards a diffractive analysis of prompt-based generative ai.
\newblock In {\em Proceedings of the CHI Conference on Human Factors in
  Computing Systems}, pages 1--15, 2024.

\bibitem{shelby2024generative}
Renee Shelby, Shalaleh Rismani, and Negar Rostamzadeh.
\newblock Generative ai in creative practice: Ml-artist folk theories of t2i
  use, harm, and harm-reduction.
\newblock In {\em Proceedings of the CHI Conference on Human Factors in
  Computing Systems}, pages 1--17, 2024.

\bibitem{srinivasan2021biases}
Ramya Srinivasan and Kanji Uchino.
\newblock Biases in generative art: A causal look from the lens of art history.
\newblock In {\em Proceedings of the 2021 ACM Conference on Fairness,
  Accountability, and Transparency}, pages 41--51, 2021.

\bibitem{mirowski2023co}
Piotr Mirowski, Kory~W Mathewson, Jaylen Pittman, and Richard Evans.
\newblock Co-writing screenplays and theatre scripts with language models:
  Evaluation by industry professionals.
\newblock In {\em Proceedings of the 2023 CHI Conference on Human Factors in
  Computing Systems}, pages 1--34, 2023.

\bibitem{chakrabarty2023creativity}
Tuhin Chakrabarty, Vishakh Padmakumar, Faeze Brahman, and Smaranda Muresan.
\newblock Creativity support in the age of large language models: An empirical
  study involving emerging writers.
\newblock {\em arXiv preprint arXiv:2309.12570}, 2023.

\bibitem{huang2023inspo}
Chieh-Yang Huang, Sanjana Gautam, Shannon~McClellan Brooks, Ya-Fang Lin, and
  Ting-Hao'Kenneth' Huang.
\newblock Inspo: Writing stories with a flock of ais and humans.
\newblock {\em arXiv preprint arXiv:2311.16521}, 2023.

\bibitem{ippolito2022creative}
Daphne Ippolito, Ann Yuan, Andy Coenen, and Sehmon Burnam.
\newblock Creative writing with an ai-powered writing assistant: Perspectives
  from professional writers.
\newblock {\em arXiv preprint arXiv:2211.05030}, 2022.

\bibitem{cherry2014quantifying}
Erin Cherry and Celine Latulipe.
\newblock Quantifying the creativity support of digital tools through the
  creativity support index.
\newblock {\em ACM Transactions on Computer-Human Interaction (TOCHI)},
  21(4):1--25, 2014.

\bibitem{shneiderman2007creativity}
Ben Shneiderman.
\newblock Creativity support tools: accelerating discovery and innovation.
\newblock {\em Communications of the ACM}, 50(12):20--32, 2007.

\bibitem{kawakami2024impact}
Reishiro Kawakami and Sukrit Venkatagiri.
\newblock The impact of generative ai on artists.
\newblock In {\em Proceedings of the 16th Conference on Creativity \&
  Cognition}, pages 79--82, 2024.

\bibitem{epstein2023art}
Ziv Epstein, Aaron Hertzmann, Investigators of~Human~Creativity, Memo Akten,
  Hany Farid, Jessica Fjeld, Morgan~R Frank, Matthew Groh, Laura Herman, Neil
  Leach, et~al.
\newblock Art and the science of generative ai.
\newblock {\em Science}, 380(6650):1110--1111, 2023.

\bibitem{jiang2023ai}
Harry~H Jiang, Lauren Brown, Jessica Cheng, Mehtab Khan, Abhishek Gupta, Deja
  Workman, Alex Hanna, Johnathan Flowers, and Timnit Gebru.
\newblock Ai art and its impact on artists.
\newblock In {\em Proceedings of the 2023 AAAI/ACM Conference on AI, Ethics,
  and Society}, pages 363--374, 2023.

\bibitem{mohamed2020decolonial}
Shakir Mohamed, Marie-Therese Png, and William Isaac.
\newblock Decolonial ai: Decolonial theory as sociotechnical foresight in
  artificial intelligence.
\newblock {\em Philosophy \& Technology}, 33:659--684, 2020.

\bibitem{zeilinger2021}
Martin Zeilinger.
\newblock Generative adversarial copy machines.
\newblock {\em Culture Machine}, 20:1--23, 2021.

\bibitem{erwin1955meaning}
Panofsky Erwin.
\newblock Meaning in the visual arts.
\newblock {\em Papers in and on Art History}, 1955.

\bibitem{hauser1951social}
Arnold Hauser.
\newblock {\em The Social History of Art Volume Four}.
\newblock 1951.

\bibitem{mirowski2024robot}
Piotr Mirowski, Juliette Love, Kory Mathewson, and Shakir Mohamed.
\newblock A robot walks into a bar: Can language models serve as creativity
  supporttools for comedy? an evaluation of llms’ humour alignment with
  comedians.
\newblock In {\em The 2024 ACM Conference on Fairness, Accountability, and
  Transparency}, pages 1622--1636, 2024.

\bibitem{gadiraju2023wouldn}
Vinitha Gadiraju, Shaun Kane, Sunipa Dev, Alex Taylor, Ding Wang, Remi Denton,
  and Robin Brewer.
\newblock " i wouldn’t say offensive but...": Disability-centered
  perspectives on large language models.
\newblock In {\em Proceedings of the 2023 ACM Conference on Fairness,
  Accountability, and Transparency}, pages 205--216, 2023.

\bibitem{mack2024they}
Kelly~Avery Mack, Rida Qadri, Remi Denton, Shaun~K. Kane, and Cynthia~L.
  Bennett.
\newblock “they only care to show us the wheelchair”: disability
  representation in text-to-image ai models.
\newblock In {\em Proceedings of the 2024 CHI Conference on Human Factors in
  Computing Systems}, CHI '24, New York, NY, USA, 2024. Association for
  Computing Machinery.

\end{thebibliography}

\end{document}